\documentclass[3p,times,twocolumn]{elsarticle}
\usepackage{graphicx}
\usepackage{lineno}
\usepackage{amsmath}
\usepackage{amssymb}
\usepackage{url}

\begin{document}

\begin{frontmatter}

\title{Why (and How) LGADs Work: Ionization, Space Charge, and Gain Saturation}

\author[address1]{N. Cartiglia\corref{mycorrespondingauthor}}
\cortext[mycorrespondingauthor]{Corresponding author}
\ead{Corresponding author. e-mail: cartiglia@to.infn.it}
\author[address4,address1]{A. R. Altamura}
\author[address1,address2]{R. Arcidiacono}
\author[address4]{M. Durando}
\author[address4]{S. Galletto}
\author[address1]{M. Ferrero}
\author[address4,address1]{L. Lanteri}
\author[address4,address1]{A. Losana}
\author[address4,address1]{L. Massaccesi}
\author[address3]{L. Menzio}
\author[address1]{F. Siviero}
\author[address4,address1]{V. Sola}
\author[address4,address1]{R. White}

\address[address1]{INFN sezione di Torino, Torino, Italy}
\address[address2]{Università del Piemonte Orientale, Novara, Italy}
\address[address3]{CERN, Geneva, Switzerland}
\address[address4]{Università degli Studi di Torino, Torino, Italy}

\begin{abstract}
The temporal resolution of Low-Gain Avalanche Detectors (LGADs), also known as Ultra-Fast Silicon Detectors (UFSDs), is governed by two contributions: jitter, arising from electronic noise and signal slew rate, and the Landau noise term, arising from the non-uniform energy deposition of minimum ionizing particles (MIPs). We show that a correct simulation of the initial ionization alone significantly overestimates the measured Landau noise. Two additional physical mechanisms are necessary to reproduce the data: space charge effects during electron/hole drift, which smooth the granularity of the initial charge distribution, and gain saturation during multiplication, which preferentially suppresses large-amplitude fluctuations. All steps of the model have been implemented in the fast simulation program Weightfield2 (WF2). The model is validated against several independent experimental observations: the evolution of the measured charge distribution with gain, the temporal resolution of events in the Landau tail, and the thickness dependence of timing performance. We also discuss a data-driven gain measurement method based on gain saturation, and implications for gain layer design.

\end{abstract}

\end{frontmatter}

\section{Introduction}
\label{sec:intro}

Low-Gain Avalanche Detectors (LGADs)~\cite{Pellegrini2014, Cartiglia2015} are silicon sensors that exploit a moderate internal gain ($\sim$10--40) to achieve excellent single-layer temporal resolution, routinely below 50~ps~\cite{Sadrozinski2018}. Their timing performance is central to several detector upgrades at the High-Luminosity LHC, including the CMS Endcap Timing Layer (ETL) and the ATLAS High-Granularity Timing Detector (HGTD)~\cite{CMS_TDR, ATLAS_TDR}.

Despite their widespread use, a complete quantitative understanding of \emph{why} LGADs achieve such good timing is still being developed. This work addresses that question from first principles, tracing the signal formation chain from the initial ionization through drift and multiplication, and identifying the physical mechanisms that govern the intrinsic temporal resolution.

All simulation results presented in this paper were obtained with the Weightfield2 (WF2) program~\cite{WF2}, a fast, publicly available simulation tool for silicon sensors available at \texttt{https://www.to.infn.it/\~{}cartigli/Weightfield2/}.

The paper is organized as follows. Section~\ref{sec:ionization} describes the MIP ionization model, including the seed distribution and long-range correlations from delta rays. Section~\ref{sec:resolution} introduces the two components of temporal resolution and provides an analytical derivation of the Landau noise term. Section~\ref{sec:discrepancy} presents the comparison between WF2 predictions and data, identifying the discrepancy that motivates two additional physical mechanisms. Section~\ref{sec:mechanisms} describes the two smoothing mechanisms --- space charge effects during drift and gain saturation during multiplication --- and introduces the phenomenological parameter $\alpha$. Section~\ref{sec:evidence} presents experimental evidence used to determine $\alpha$. Section~\ref{sec:combined} shows that, with $\alpha$ fixed, WF2 predictions agree with measured temporal resolution across all sensor thicknesses. Section~\ref{sec:why} summarises why LGADs achieve good timing through the interplay of these three mechanisms. Section~\ref{sec:gain_measurement} introduces a novel data-driven gain measurement method. Section~\ref{sec:design} discusses implications for gain layer design. Section~\ref{sec:conclusions} summarizes the conclusions.

\section{MIP Ionization in Silicon Sensors}
\label{sec:ionization}

\subsection{The Seed Distribution Model}

The energy deposition of a MIP in a silicon sensor of arbitrary thickness is modeled in WF2 as a sum of local deposits drawn independently from a \emph{seed distribution} --- the energy deposited in a 1-$\mu$m thick silicon layer. This seed distribution has two physically distinct components:

\begin{itemize}
  \item A soft scattering component, corresponding to small, frequent energy transfers. Because the cross-section diverges at low $T$, the vast majority of interactions transfer very little energy. In WF2, these are modeled as a Gaussian contribution with mean $\mu_\text{soft}(d)$ and standard deviation $\sigma_\text{soft}$ per micron slice.
  \item A hard scattering component, corresponding to more energetic collisions sampled from the $1/T^2$ distribution over $[T_\text{min}, T_\text{max}]$, that produce $\delta$-rays --- knock-on electrons carrying significant kinetic energy away from the primary track. The value of $T_\text{min}$ sets the width of this energy range and hence the normalisation of the distribution. A $\delta$-ray with $T \sim 30$--$100$~keV can travel tens of microns from the primary track. These events populate the high-energy Landau tail.
\end{itemize}

\begin{figure}[t!]
  \vspace{.2cm}
  \centering
  \includegraphics[width=.70\linewidth]{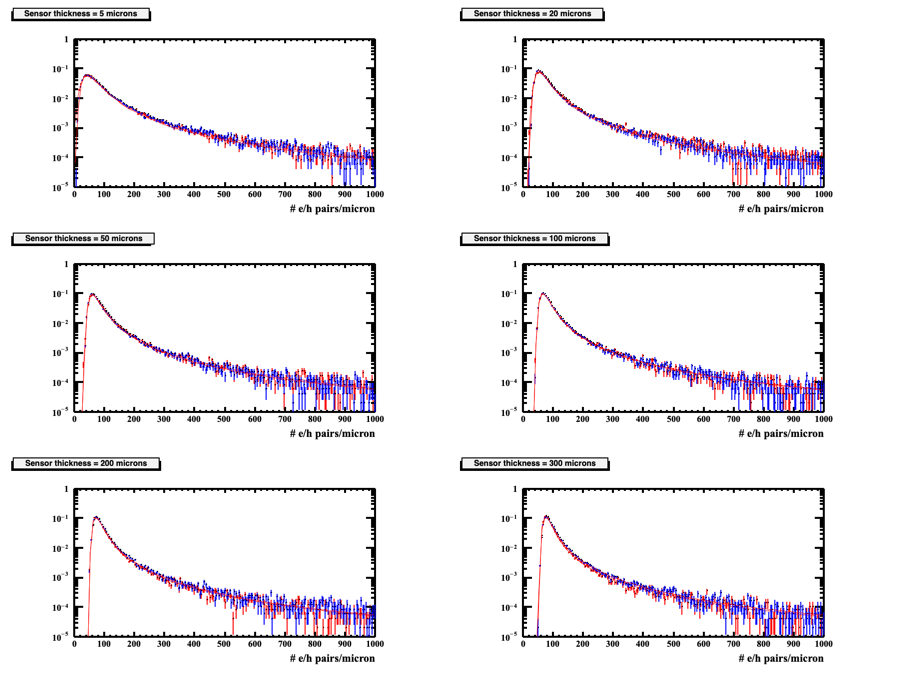}
  \caption{Measured (red) and WF2-simulated Landau energy distributions (blue). The simulation correctly reproduces both the MPV and the FWHM as a function of sensor thickness.}
  \label{fig:landau_validation}
\end{figure}

Concretely, in each 1-$\mu$m slice the number of hard collisions $N$ is drawn from a Poisson distribution with mean $\nu \cdot \Delta x$, where $\nu$ is the mean collision rate per micron. Each collision energy is sampled from the truncated $1/T^2$ distribution using the exact inverse-CDF method:
\begin{equation}
  T = \frac{1}{\,1/T_\text{min} - u\,(1/T_\text{min} - 1/T_\text{max})\,}, \quad u \sim \mathcal{U}(0,1)
  \label{eq:inverseCDF}
\end{equation}
with $T_\text{min} = 26$~eV and $T_\text{max} = 600$~keV. The results are insensitive to $T_\text{max}$ for values above $\sim 200$~keV, as the $1/T^2$ spectrum strongly suppresses large energy transfers. The soft component is added as an independent Gaussian draw ($\mu_\text{soft}$, $\sigma_\text{soft} = 10$~eV) per micron. The total slice energy (hard + soft) is converted to electron-hole pairs using the mean ionisation energy in silicon, $\epsilon = 3.6$~eV/pair. Both $\nu$ and $\mu_\text{soft}$ depend on sensor thickness following empirical scaling relations:
\begin{equation}
  \nu(d) = \frac{3.8}{d^{0.43}}, \qquad \mu_\text{soft}(d) = 22\ln(d) - 30 \text{ [eV]}
  \label{eq:scaling}
\end{equation}
This thickness dependence reflects the fact that the relative weight of the hard and soft components in the energy deposition changes systematically with thickness. The scaling functions were determined by fitting the measured MPV and FWHM of the Landau distribution across sensor thicknesses from 5 to 300~$\mu$m. The total energy deposition in a sensor of thickness $d$ is obtained by drawing a sequence of $d/1\,\mu\text{m}$ random samples from the seed distribution and summing them. Using this approach, WF2 correctly reproduces the measured MPV~\cite{Meroli2011} and FWHM of the Landau distribution as a function of sensor thickness (Figure~\ref{fig:landau_validation}), confirming the validity of the ionization model.

\subsection{Long-Range Correlations from Delta Rays}

In WF2, each $\delta$-ray of energy $T$ is assigned a range using the empirical CSDA relation for electrons in silicon~\cite{Leo1994}:
\begin{equation}
  R\,[\mu\text{m}] = 0.04 \cdot \left(\frac{T}{1\,\text{keV}}\right)^{1.75}
  \label{eq:deltarange}
\end{equation}
$\delta$-rays with $R < 2\,\mu$m deposit their energy locally at the point of creation. For $R \geq 2\,\mu$m, the energy is distributed along the track using a power-law stopping profile derived by inverting Eq.~\eqref{eq:deltarange} ($R \propto T^{1.75}$ implies $T \propto R^{1/1.75}$):
\begin{equation}
  E_\text{remaining}(x) = T \cdot \left(\frac{R - x}{R}\right)^{1/1.75}
\end{equation}
so that the energy deposited in each 1-$\mu$m slice is the difference in remaining energy between its start and end. This reflects the physical behaviour of electrons, whose energy loss per unit length increases as they slow down, so that the deposited energy per slice rises toward the end of the range. If the range of a $\delta$-ray exceeds the remaining distance to the sensor edge, energy deposition is truncated at the boundary and the remaining energy escapes the sensor undetected.

\section{Components of the Temporal Resolution}
\label{sec:resolution}

\subsection{Jitter and Landau Noise}

The temporal resolution $\sigma_t$ of an LGAD is decomposed into two independent contributions~\cite{Cartiglia2017}:
\begin{equation}
  \sigma_t^2 = \sigma_\text{jitter}^2 + \sigma_\text{Landau}^2
\end{equation}

Jitter arises from electronic noise $\sigma_V$ and the signal slew rate $dV/dt$ at threshold crossing:
\begin{equation}
  \sigma_\text{jitter} = \frac{\sigma_V}{dV/dt}
\end{equation}
It dominates at low gain, where the signal amplitude is small relative to the noise.

Landau noise, also called the non-uniform ionization term, arises from the event-to-event variability of the charge deposition. Each MIP deposits charge in a different spatial pattern along the track, producing a different signal shape and hence a different trigger time. This term dominates at high gain, where jitter is negligible.

Throughout this work, the trigger time is defined by a Constant Fraction Discriminator (CFD) at 30\% of the signal amplitude, both in the experimental data and in the WF2 simulation.

\subsection{Analytical Derivation of the Landau Noise Term}
\label{sec:analytical}

Following Riegler~\cite{Riegler2025}, we derive the general features of the Landau noise term.   Consider a sensor of thickness $d$ in which a single electron-hole pair is created at a random, uniformly distributed position $x \in [0, d]$. The electron drifts to the collecting electrode with velocity $v$, arriving at time $t = x/v$. Since $x$ is uniform on $[0, d]$, the standard deviation of the arrival time is:
\begin{equation}
  \sigma_0(d) = \frac{d}{v\sqrt{12}}
\end{equation}
This is the fundamental upper bound on the Landau noise: it corresponds to total ignorance of the charge creation position. For a laser that deposits $N$ electron-hole pairs uniformly along the track, the uncertainty on the trigger time improves as $1/\sqrt{N}$ (by analogy with the mean of $N$ uniform random variables). 
\begin{equation}
 \sigma^{\text{Laser}} = \frac{\sqrt{d}}{v\sqrt{12n}}
 \label{eq:sigma_laser}
\end{equation}
where $n$ is the linear density of electron-hole pairs (pairs per unit length). For the MIP case, the charge deposition is non-uniform and the analytical treatment becomes significantly more complex; the full derivation is given in~\cite{Riegler2025}.

\begin{figure}[t!]
 \vspace{.2cm}
 \centering
 \includegraphics[width=.70\linewidth]{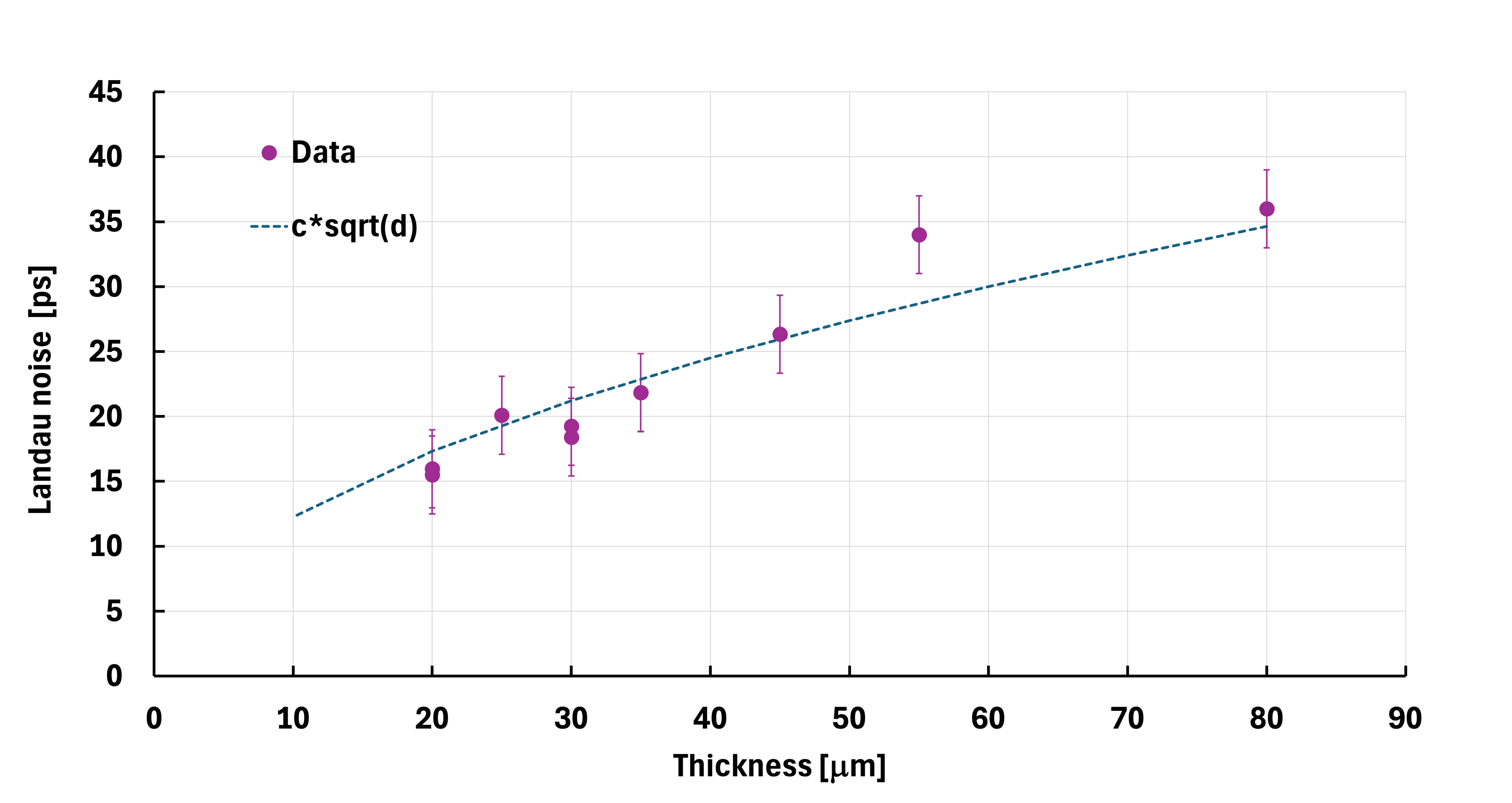}
 \caption{Measured values of the Landau noise as a function of sensor thickness and a line that shows the $\propto\sqrt{d}$ behaviour.}
 \label{fig:landau_noise}
\end{figure}

Equation~\eqref{eq:sigma_laser}  reveals the fundamental structure of the Landau noise:  it grows with the square root of the thickness and decreases with drift velocity. Operating at fields sufficient to saturate the drift velocity $v$, and reducing sensor thickness, are the most effective ways to reduce this term. Specifically, for the MIP case, the $N$ charge carriers are not uniformly distributed but follow the non-uniform Landau deposition. The resulting Landau noise is larger than for a laser, and its precise value depends on the specific ionization pattern in each event ~\cite{Riegler2025}.  Figure~\ref{fig:landau_noise} shows  that the theoretical $\sqrt{d}$ scaling well matches the measured values of Landau noise  ~\cite{Siviero2022}.

\section{The Discrepancy Between Simulation and Data}
\label{sec:discrepancy}

\begin{figure}[t!]
  \vspace{.2cm}
  \centering
  \includegraphics[width=.75\linewidth]{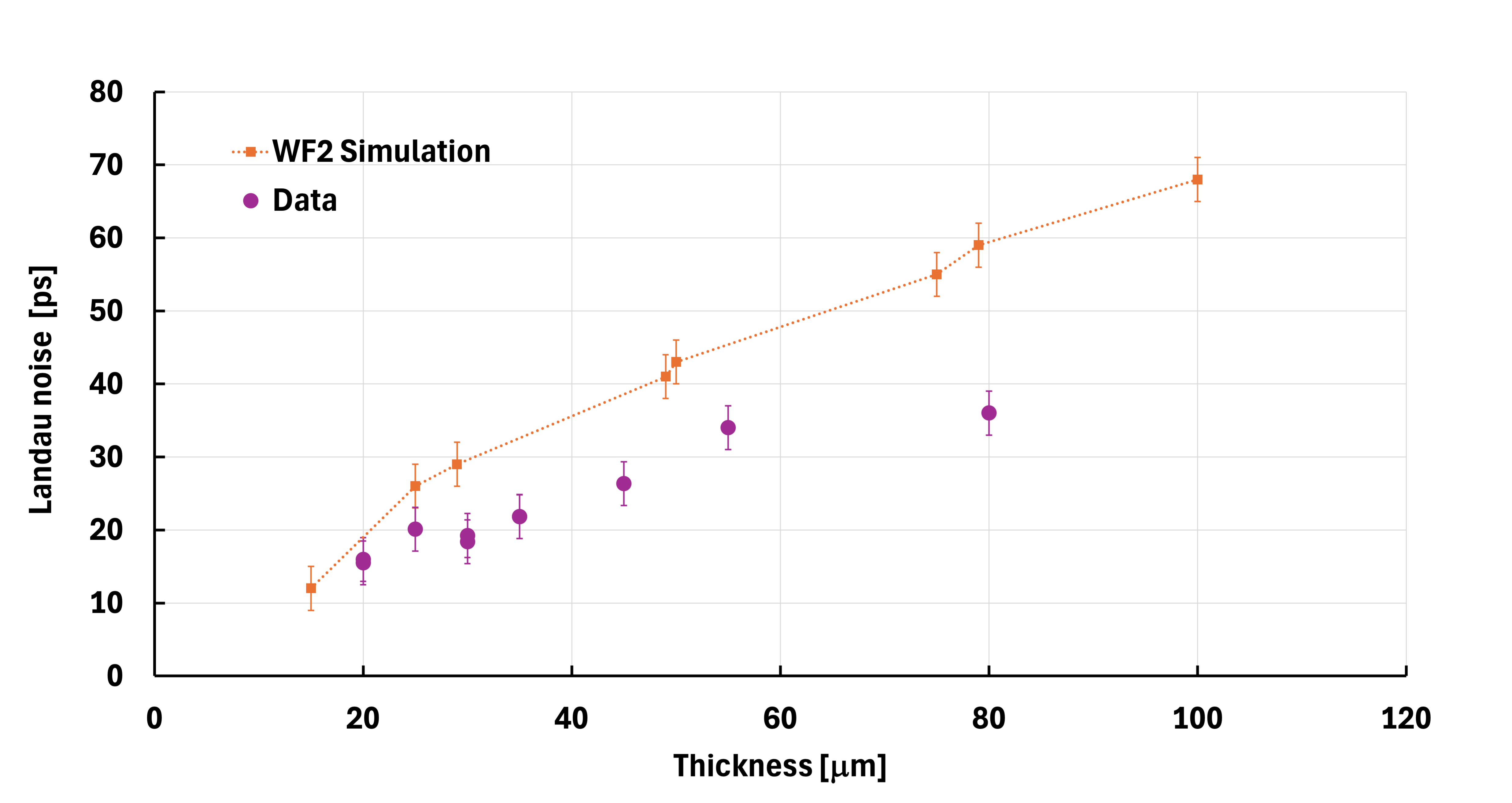}
  \caption{Temporal resolution (Landau noise contribution, jitter subtracted) vs. sensor thickness. The WF2 prediction using only the initial Landau ionization significantly overestimates the measured values, especially for thick sensors.}
 \label{fig:res_discrepancy}
\end{figure}

Figure~\ref{fig:res_discrepancy} compares the measured temporal resolution (jitter subtracted, Landau noise contribution only) with the WF2 predictions obtained using the Landau ionization model described above. Two key observations emerge:
\begin{enumerate}
  \item The simulated Landau noise is considerably larger than the measured one across all sensor thicknesses.
  \item The discrepancy grows with increasing sensor thickness.
\end{enumerate}

This discrepancy implies that the actual LGAD signal shape is smoother than the ionization model implemented  in WF2. Two physical mechanisms, not included in the simulation, are responsible for this discrepancy: space charge effects during drift and gain saturation during multiplication.

\section{Smoothing Mechanisms}
\label{sec:mechanisms}

\subsection{Space Charge Effects During Drift}

\begin{figure}[t!]
  \vspace{.2cm}
  \centering
  \includegraphics[width=.80\linewidth]{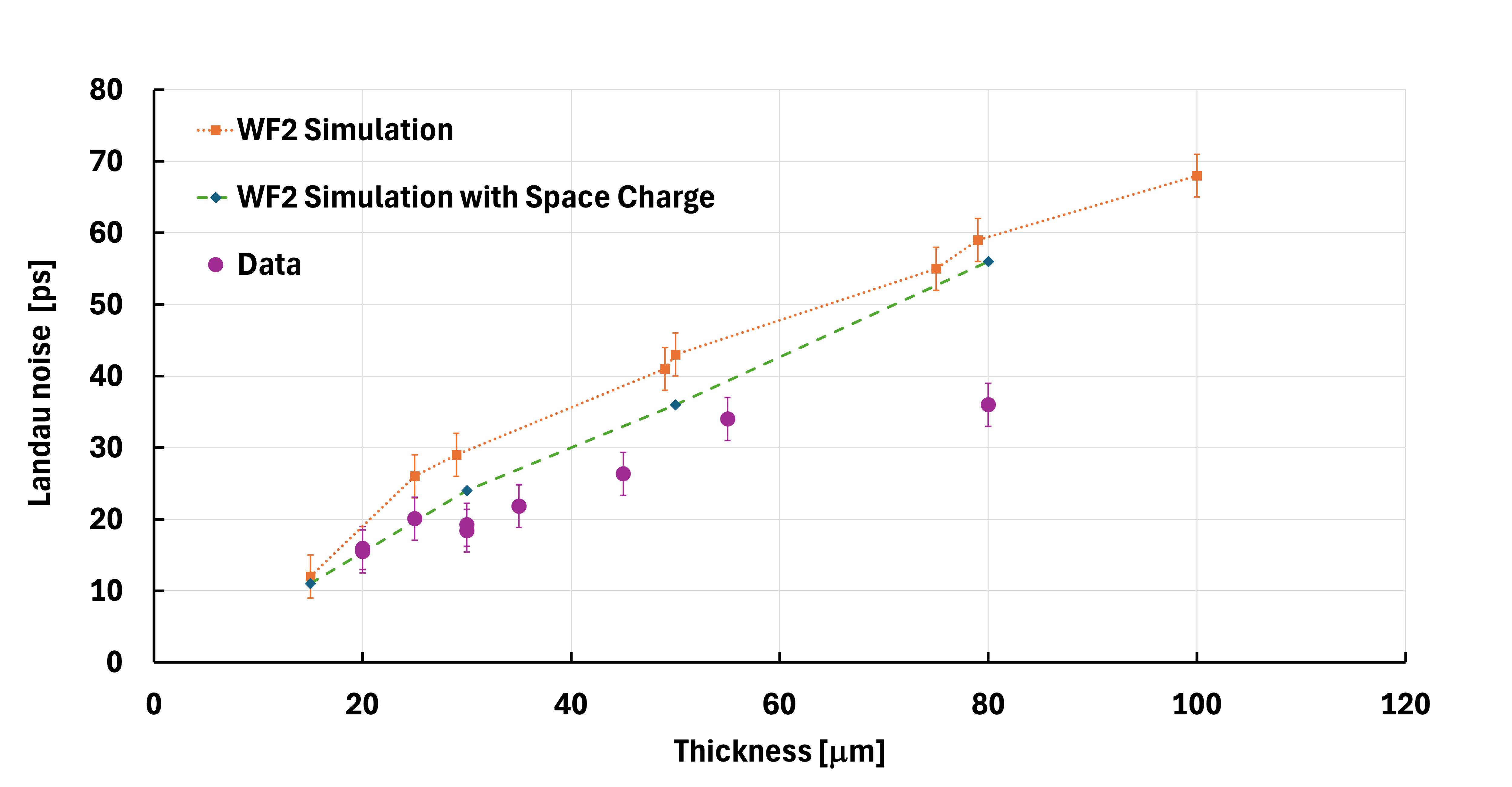}
  \caption{The measured Landau noise contribution is compared with the WF2 simulation with and without space charge effects.}
  \label{fig:space_charge}
\end{figure}

During the drift of electrons and holes toward the electrodes, the charge clusters interact via their mutual Coulomb repulsion. This is distinct from diffusion: high charge-density clusters experience a stronger repulsive force and expand more rapidly than low-density clusters. The net effect is that the initially granular charge distribution becomes smoother during drift. In WF2, this is modeled as a one-dimensional electric field generated by the clusters at greater and smaller depth along the drift direction relative to a given cluster, which modifies the local drift velocity. After the drift, the electron cloud is elongated and has lower peak density. As Figure~\ref{fig:space_charge} shows,  including space charge effects improves agreement with data slightly but is insufficient alone to explain the full discrepancy.

\subsection{Gain Saturation}
\label{sec:saturation}

The dominant smoothing mechanism is gain saturation. In the gain volume, as the electrons generated by the multiplication drift toward the n$^+$ cathode and the holes drift toward the gain implant, a field opposing the LGAD bias field is created, locally reducing the effective multiplication. At high gain or charge density, the field created by the e/h carriers lowers the bias field considerably, reducing the gain value. As a result, large charge deposits are amplified less than small ones. Gain saturation,  therefore,  acts as a nonlinear compressor, selectively suppressing the large-amplitude fluctuations that most degrade timing. In WF2, gain saturation is implemented as follows. At each simulation step, the total number of electron-hole pairs present inside the gain layer is counted. All electrons are placed at the n$^+$ junction and all holes at the gain implant, effectively modelling the charge as two thin sheets deposited over an area of 1~$\mu$m$^2$, consistent with the transverse extent of the primary ionization column of a MIP. The electric field generated by this charge configuration is:
\begin{equation}
  E_\text{sat} = \alpha \cdot \frac{N_{eh} \cdot e}{\epsilon_\text{Si} \cdot A}
\end{equation}
where $N_{eh}$ is the number of electron-hole pairs in the gain layer, $e$ is the elementary charge, $\epsilon_\text{Si}$ is the permittivity of silicon, $A = 1\,\mu\text{m}^2$ is the assumed charge area, and $\alpha$ is a phenomenological correction factor. The simplified model places all charges at fixed positions, whereas in reality the charge distribution is dynamic and spatially extended, so the actual field is different from the static estimate; $\alpha$ compensates for this simplification. The field $E_\text{sat}$ opposes the bias field and is subtracted from the total electric field (bias + gain implant) in the gain layer. The reduced effective field lowers the impact ionisation coefficient, and hence the local gain. Large charge deposits generate a stronger opposing field, are amplified less, and the gain saturation effect is therefore stronger at higher incoming charge and at higher bias voltage.

\begin{figure}[t!]
  \vspace{.2cm}
  \centering
  \includegraphics[width=.70\linewidth]{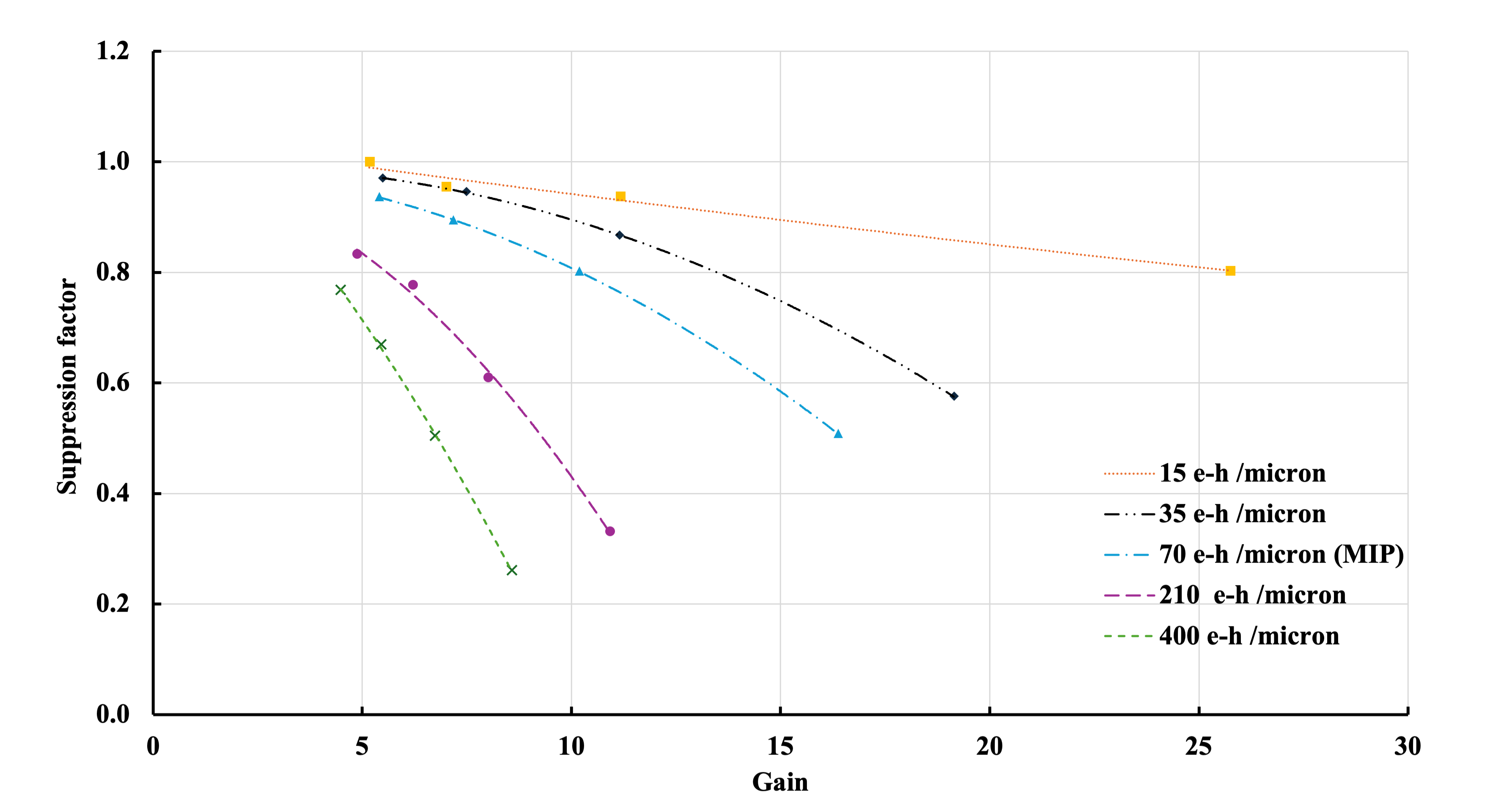}
  \caption{Simulated gain suppression factor as a function of gain. Gain saturation is stronger for larger charge deposits and higher gains.}
  \label{fig:gain_quench}
\end{figure}

Figure~\ref{fig:gain_quench} shows the simulated gain suppression factor as a function of gain, demonstrating the characteristic suppression at high charge density and high gain.

\section{Experimental Evidence for Gain Saturation}
\label{sec:evidence}

The value of $\alpha$ introduced in Section~\ref{sec:saturation} is not a free parameter: it is determined by fitting WF2 to the experimental observations described in this section. Once fixed, the same value of $\alpha$ is used in all subsequent simulations without further adjustment.

\subsection{Evolution of the Charge Distribution with Gain}

\begin{figure}[t!]
  \vspace{.2cm}
  \centering
  \includegraphics[width=.90\linewidth]{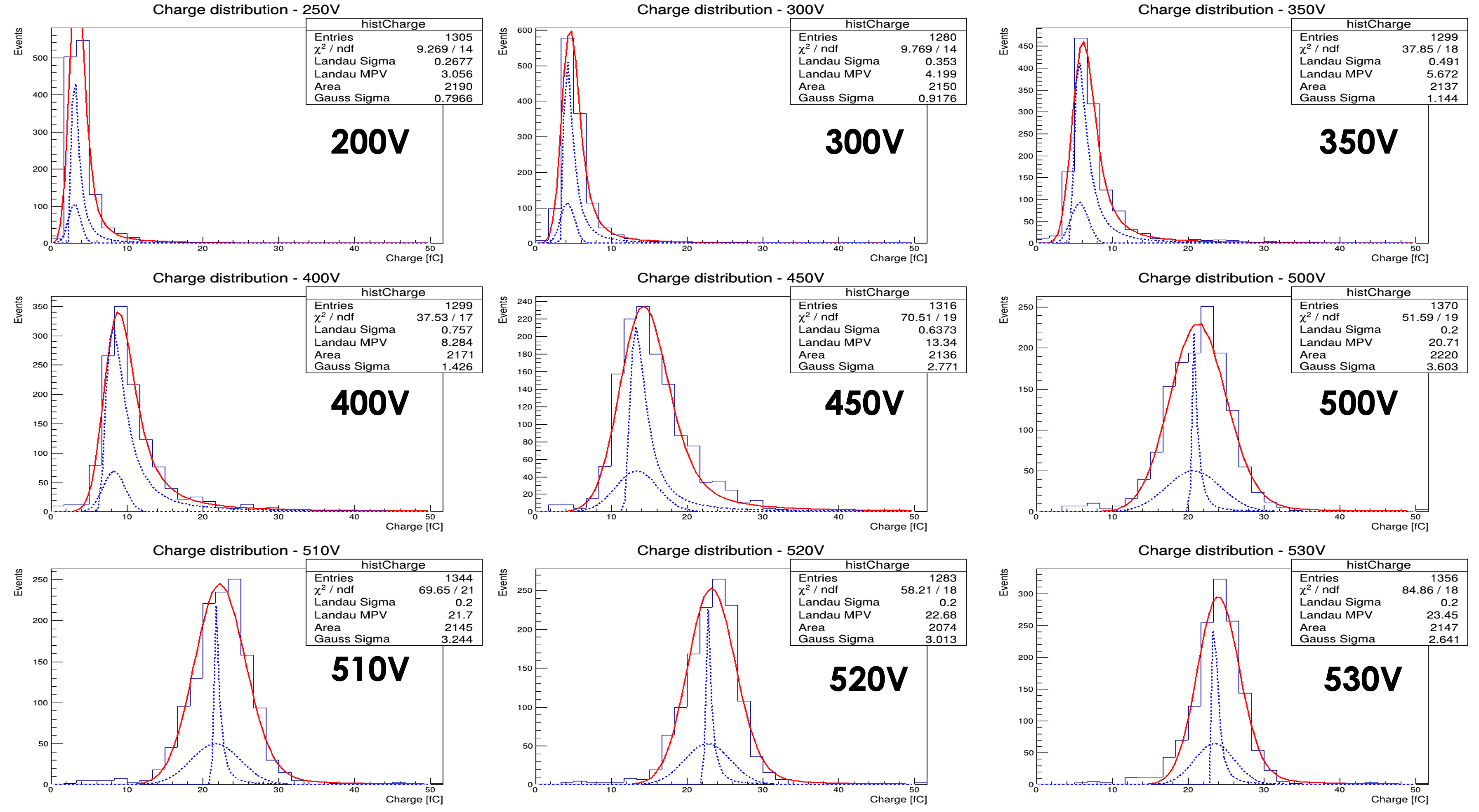}
  \caption{Measured charge distribution for a 50-$\mu$m LGAD at bias voltages from 200~V to 530~V (corresponding to gain values from $\sim$5 to $\sim$40). The distribution evolves from Landau-like at low gain to approximately Gaussian at high gain, a direct signature of gain saturation.}
  \label{fig:landau_evolution}
\end{figure}

Since gain saturation preferentially suppresses large charge deposits, the measured amplitude distribution of an LGAD is distorted relative to the initial Landau, and the distortion increases with gain. At low gain ($\sim$5), the measured distribution matches the standard Landau parametrization for PIN diodes~\cite{Meroli2011}. As the gain increases, the high-amplitude tail is progressively suppressed, and at very high gain the distribution becomes approximately Gaussian (Figure~\ref{fig:landau_evolution}).

\begin{figure}[t!]
  \vspace{.2cm}
  \centering
  \includegraphics[width=.68\linewidth]{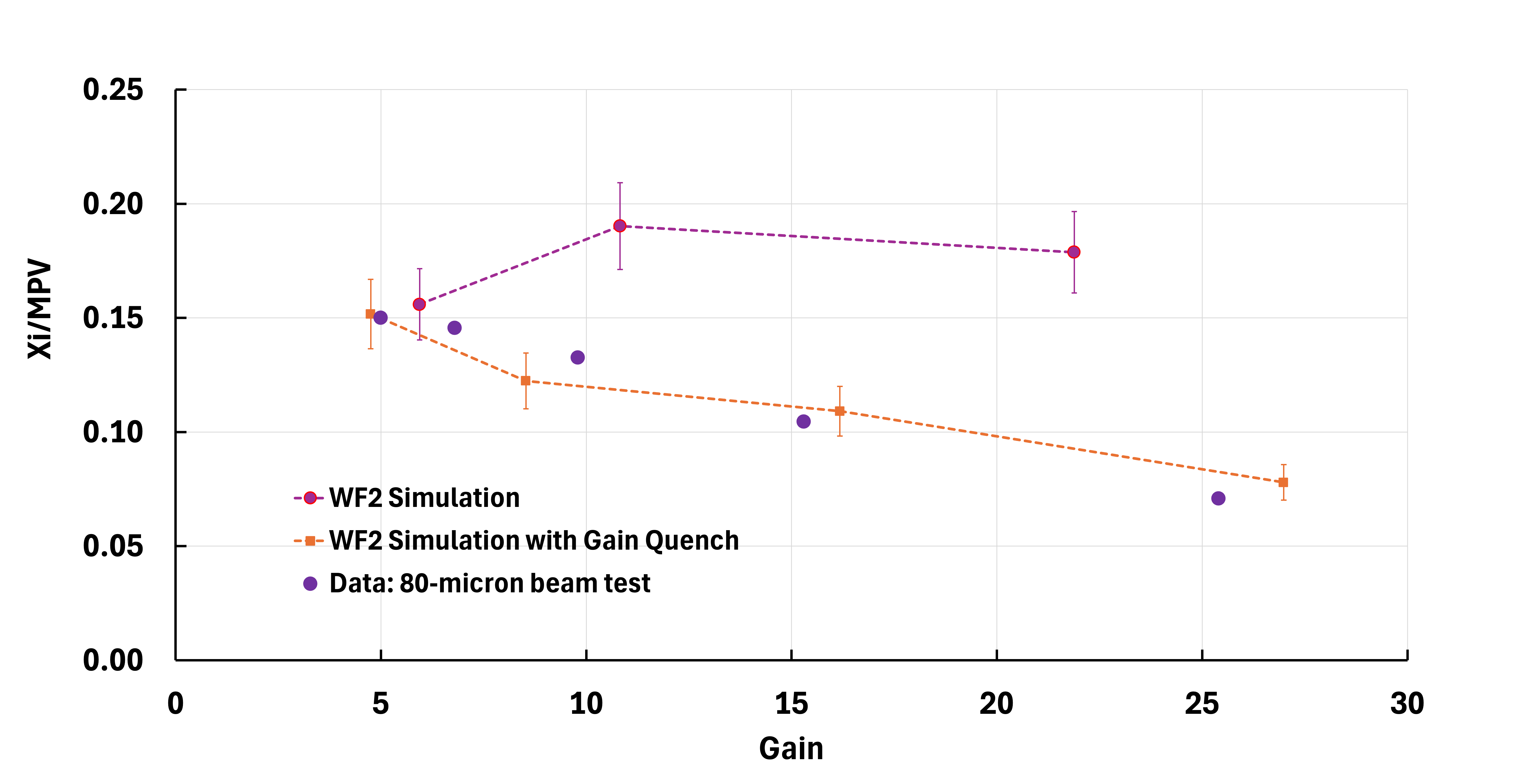}
  \caption{$\xi/\text{MPV}$ vs. gain for data (80-$\mu$m beam test), WF2 without gain saturation, and WF2 with gain saturation. Gain saturation is essential to reproduce the measured trend.}
  \label{fig:sigma_mpv}
\end{figure}

This evolution of the charge distribution is quantified by the ratio $\xi/\text{MPV}$, where $\xi$ is the width parameter of the fitted Landau function and MPV is its most probable value. Figure~\ref{fig:sigma_mpv} shows that WF2 with gain saturation correctly reproduces the measured decrease of $\xi/\text{MPV}$ with gain, while WF2 without gain saturation predicts an approximately constant ratio --- in clear disagreement with data.

\subsection{Temporal Resolution as a Function of Landau Position}

\begin{figure}[t!]
  \vspace{.2cm}
  \centering
  \includegraphics[width=1.\linewidth]{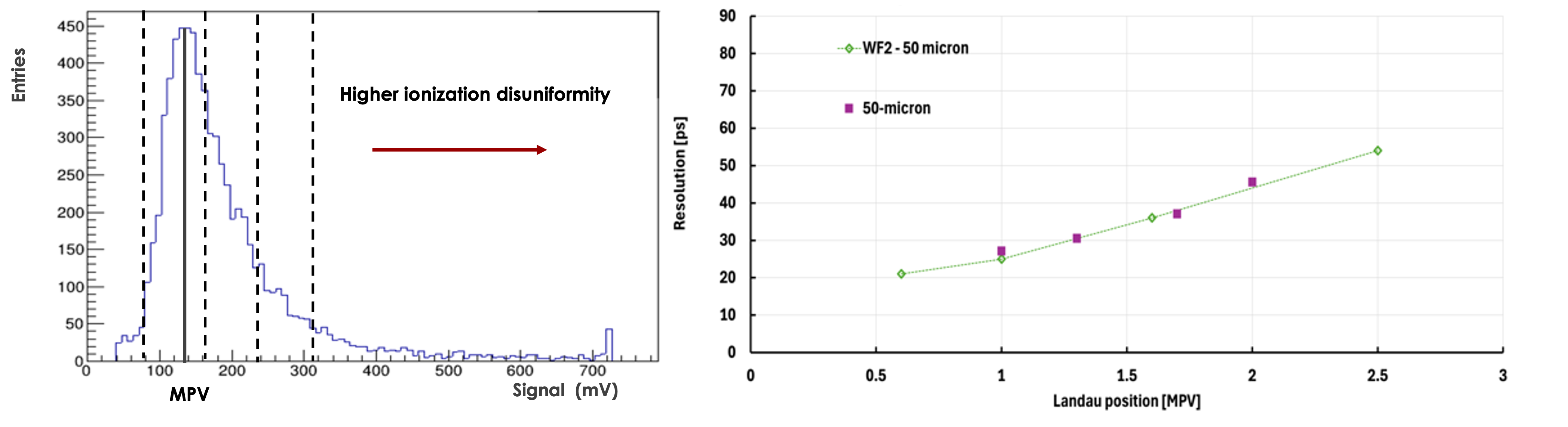}
  \caption{Temporal resolution as a function of event position in the Landau distribution. Events in the high-amplitude tail have significantly worse timing. The agreement between WF2 (including gain saturation) and data confirms the model.}
  \label{fig:res_vs_landau}
\end{figure}

Events in the high-amplitude tail of the Landau distribution contain large, non-uniform energy deposits that produce irregular signal shapes and are expected to have worse temporal resolution. This is confirmed experimentally~\cite{Siviero2022}: the timing resolution degrades progressively for events further into the Landau tail (Figure~\ref{fig:res_vs_landau}). The agreement between WF2 and data confirms that the simulation correctly captures both the initial ionization and the gain saturation. The degradation is more pronounced in thicker sensors, consistently with the higher probability of large deposits.

\section{Combined Effect and Agreement with Data}
\label{sec:combined}

\begin{figure}[t!]
  \vspace{.2cm}
  \centering
  \includegraphics[width=.75\linewidth]{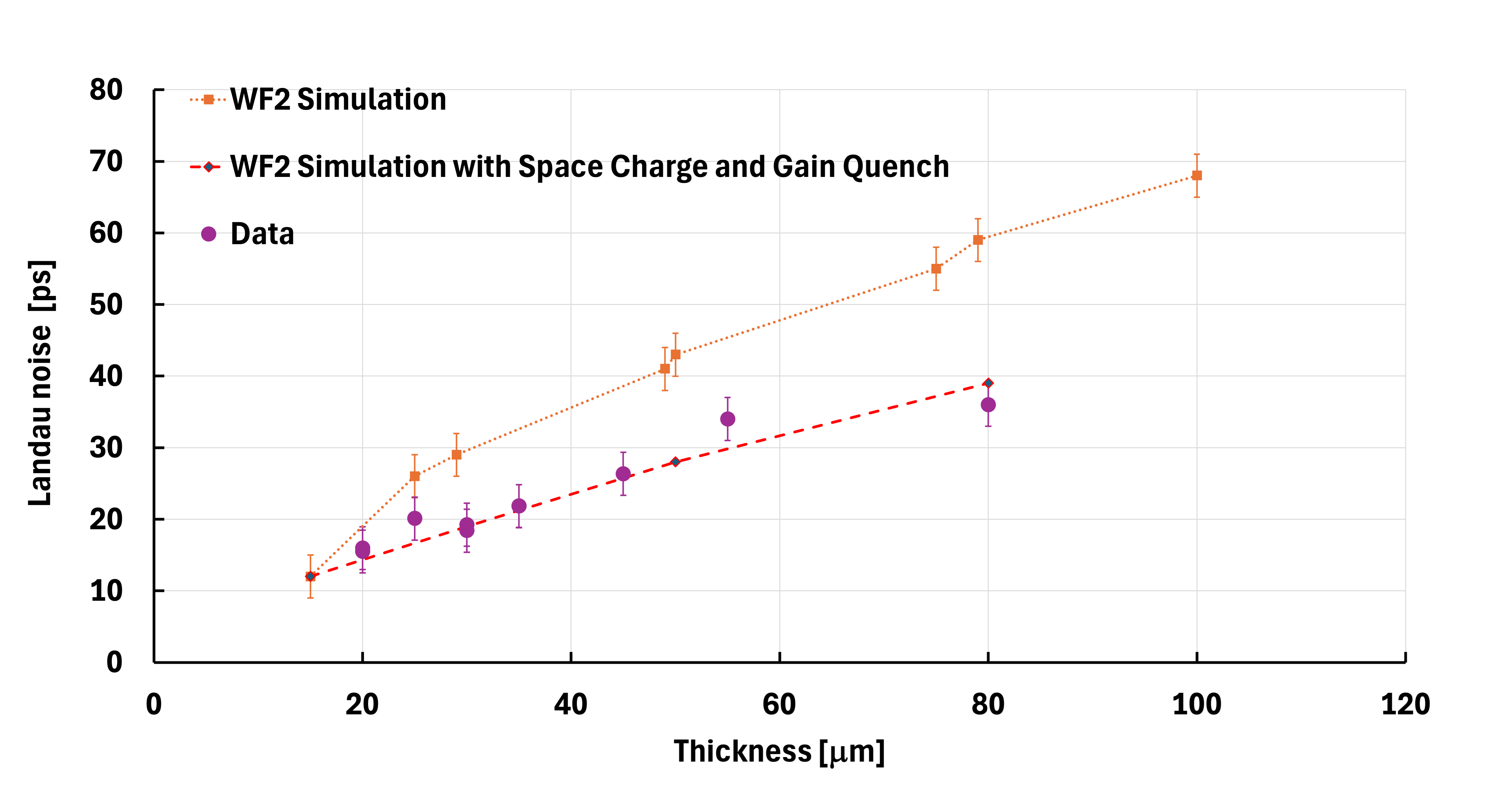}
  \caption{Temporal resolution vs. sensor thickness. WF2 predictions including space charge effects and gain saturation agree well with data. The gain saturation contribution is the dominant effect.}
  \label{fig:res_full}
\end{figure}

With $\alpha$ fixed by the measurements in Section~\ref{sec:evidence}, no further free parameters remain. Including both space charge effects and gain saturation in WF2 brings the predicted temporal resolution into good agreement with data across all sensor thicknesses (Figure~\ref{fig:res_full}). Gain saturation dominates the correction; space charge effects provide a smaller but non-negligible contribution.

\section{Why LGADs Work}
\label{sec:why}

\begin{figure}[t!]
  \vspace{.2cm}
  \centering
  \includegraphics[width=.75\linewidth]{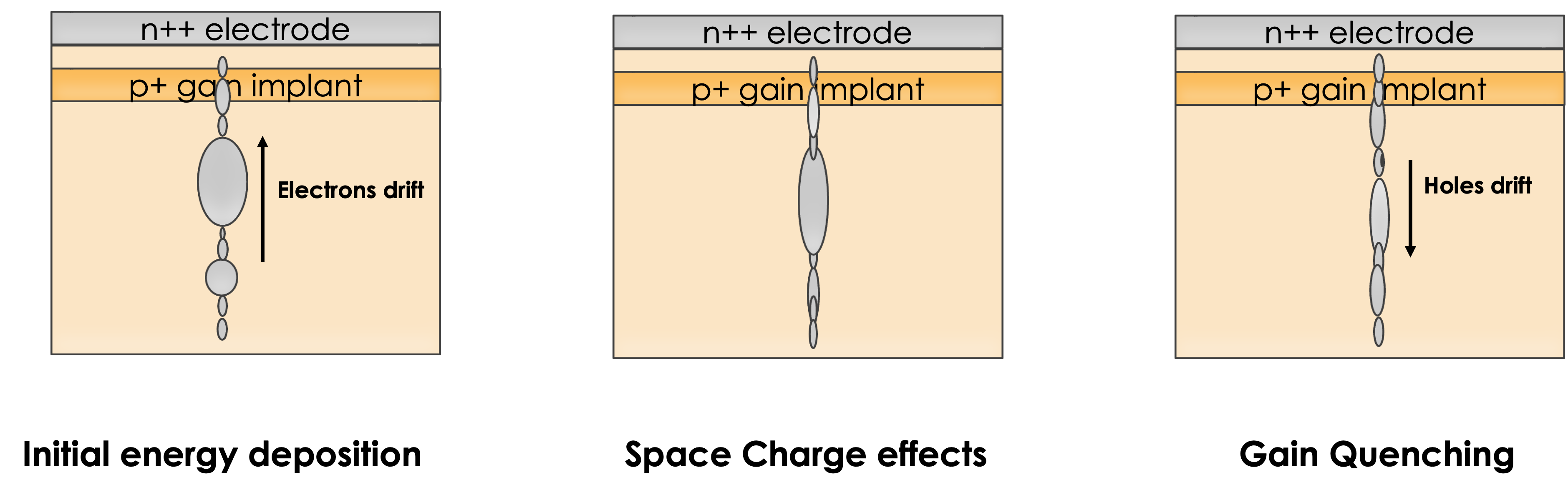}
  \caption{Sketch of the 3 steps leading to the LGAD temporal resolution: initial ionization, space charge, and gain saturation.}
  \label{fig:summary}
\end{figure}

The LGAD intrinsic temporal resolution is governed by three sequential mechanisms, sketched in Figure~\ref{fig:summary}:
\begin{enumerate}
  \item Initial non-uniform ionization: the MIP deposits charge non-uniformly along the track. 
  \item Space charge smoothing during drift: Coulomb repulsion between clusters partially smears out the initial charge non-uniformity.
  \item Gain saturation during multiplication: large deposits are amplified less, compressing the amplitude fluctuations and regularizing the signal.
\end{enumerate}

LGADs work because effects (2) and (3) progressively reduce the impact of the initial Landau variability, bringing the effective Landau noise well below what the raw ionization statistics would predict.

\section{Data-Driven Gain Measurement via Landau Distortion}
\label{sec:gain_measurement}

\begin{figure}[t!]
  \vspace{.2cm}
  \centering
  \includegraphics[width=1.\linewidth]{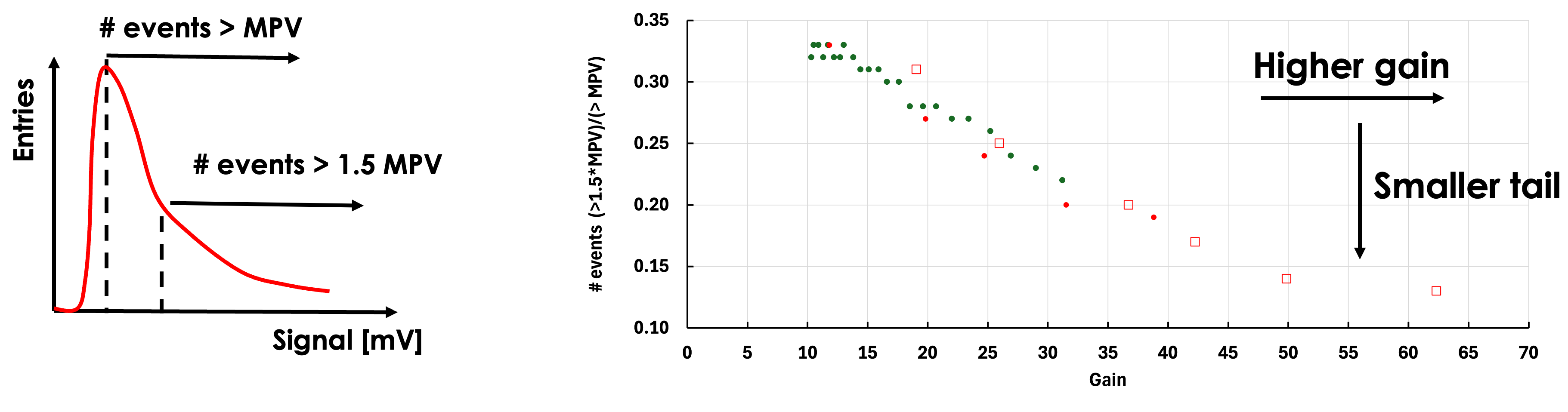}
  \caption{Landau Tail Fraction (LTF) as a function of gain for a 50~$\mu$m thick sensor. The LTF decreases monotonically with gain, providing a data-driven method to evaluate the gain.}
 \label{fig:Ratio}
\end{figure}

The progressive suppression of the Landau tail by gain saturation provides a novel data-driven method for gain measurement. The observable is the Landau Tail Fraction (LTF), defined as $(N_{>1.5\,\text{MPV}})/(N_{>\text{MPV}})$: at higher gain, stronger saturation suppresses the tail more, so the LTF decreases monotonically with gain. The threshold of $1.5\times\text{MPV}$ is chosen as a compromise: it is sufficiently far into the tail to be sensitive to gain saturation, while retaining enough statistics to be measured reliably. The method works as follows: using a set of measurements in which the gain is also known independently (e.g.\ from charge calibration), the LTF-vs-gain curve is established empirically. Subsequently, the gain of any new measurement can be extracted from its LTF alone.
The LTF is preferable to the ratio $\xi/\text{MPV}$, which is more difficult to compute correctly because the low-amplitude part of the Landau distribution is often distorted by system noise, especially at low gain. The LTF is more robust, as it depends only on the well-measured high-amplitude region. Figure~\ref{fig:Ratio} shows how the LTF decreases as a function of gain for a 50~$\mu$m thick sensor.

\section{Implications for Gain Layer Design}
\label{sec:design}

Since gain saturation is the dominant mechanism suppressing Landau noise, an ideal gain layer should maximize saturation: stronger saturation compresses the amplitude fluctuations arising from Landau variability, directly reducing the Landau noise and improving timing resolution. A deeper gain implant operates at a lower electric field (at the same gain) compared to a shallow one. At lower fields, the ionization mean free path $\lambda$ has a steeper dependence on the field ($d\lambda/dE$ is larger), so a given field perturbation from the drifting charge has a larger relative effect on $\lambda$ and hence on the local gain, leading to stronger saturation.

\begin{figure}[t!]
  \vspace{.2cm}
  \centering
  \includegraphics[width=.68\linewidth]{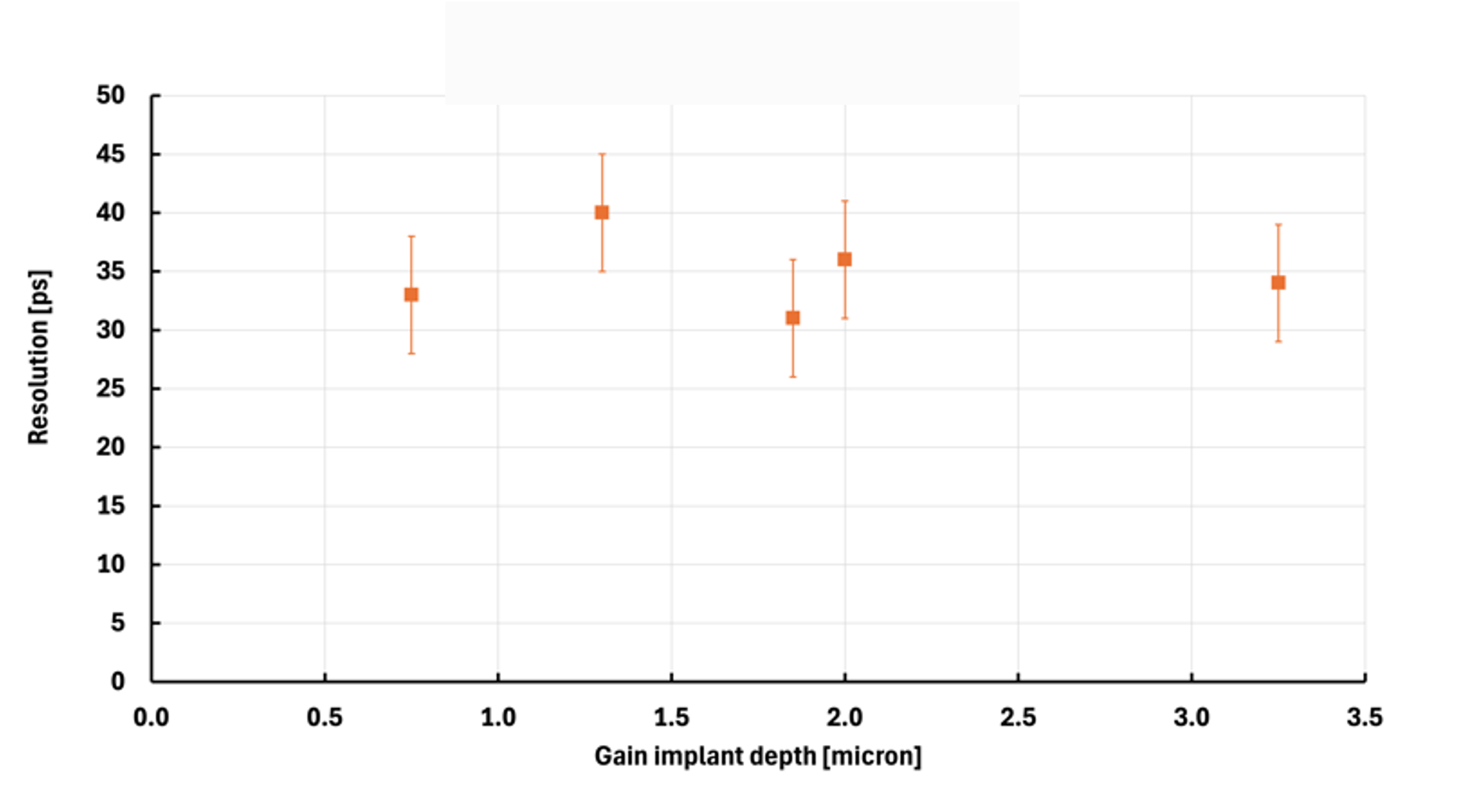}
  \caption{WF2-predicted temporal resolution (events with amplitude $>2\times\text{MPV}$, 80-$\mu$m sensor, gain $\sim$15, jitter = 0) vs. gain implant depth. No significant dependence is observed.}
  \label{fig:gain_implant}
\end{figure}

However, WF2 simulations predict that in the typical LGAD operating regime (gain $\sim$15), varying the gain implant depth does not significantly improve the Landau noise contribution to temporal resolution (Figure~\ref{fig:gain_implant}). The simulation focuses on events with amplitude $>2\times\text{MPV}$, as these high-charge events are the most granular and therefore the most sensitive to any improvement or degradation due to gain saturation. Straightforward modifications of the gain layer geometry are unlikely to yield significant timing improvements through enhanced saturation in the standard operating regime. Designing intrinsically higher-saturating gain layers --- through novel implant profiles, material engineering, or operating conditions outside the standard regime --- would lead to a lower Landau noise and better overall temporal resolution. 

\section{Conclusions}
\label{sec:conclusions}

We have presented a detailed model for the intrinsic temporal resolution of LGADs, implemented in the Weightfield2 simulation program and validated against a comprehensive set of experimental data. The main conclusions are:

\begin{itemize}
  \item The temporal resolution has two contributions: jitter and Landau noise. The analytical derivation shows that the Landau noise scales as $\sqrt{d}/v$, where $d$ is the sensor thickness and $v$ is the drift velocity.
  \item Simulating only the initial Landau ionization considerably overestimates the Landau noise. Two additional mechanisms are essential: space charge effects and, above all, gain saturation.
  \item Gain saturation transforms the initial Landau charge distribution progressively into a Gaussian as gain increases, preferentially suppressing the large-amplitude fluctuations that degrade timing.
  \item A single value of $\alpha$ simultaneously accounts for three apparently independent experimental observations: the evolution of $\xi/\text{MPV}$ with gain, the degradation of temporal resolution for events in the Landau tail, and the measured LGAD temporal resolution as a function of sensor thickness. The consistency of these three fits provides strong evidence that the gain saturation model captures the underlying physics.
  \item The Landau Tail Fraction (LTF), defined as $(N_{>1.5\,\text{MPV}})/(N_{>\text{MPV}})$, provides a new data-driven method for LGAD gain measurement that requires no simulation and relies only on the well-measured high-amplitude region of the charge distribution.
  \item In the standard operating regime, gain implant depth does not significantly reduce Landau noise; designing higher-saturating gain layers remains an open problem.
\end{itemize}

In summary: LGADs work because the initial non-uniform ionization is progressively smoothed by space charge effects and, predominantly, by gain saturation. Understanding this chain is the key to designing next-generation timing detectors with improved performance.

{}


\begin{thebibliography}{}

\bibitem{Pellegrini2014}
  G. Pellegrini et al.,
  ``Technology developments and first measurements of Low Gain Avalanche Detectors (LGAD) for high energy physics'',
  Nucl. Instrum. Meth. A \textbf{765} (2014) 12.

\bibitem{Cartiglia2015}
  N. Cartiglia et al.,
  ``Design optimization of Ultra-Fast Silicon Detectors'',
  Nucl. Instrum. Meth. A \textbf{796} (2015) 141.

\bibitem{Cartiglia2017}
  N. Cartiglia et al.,
  ``Beam test results of a 16~ps timing system based on ultra-fast silicon detectors'',
  Nucl. Instrum. Meth. A \textbf{850} (2017) 83.

\bibitem{Sadrozinski2018}
  H. Sadrozinski, A. Seiden, N. Cartiglia,
  ``4-Dimensional tracking with Ultra-Fast Silicon Detectors'',
  Rep. Prog. Phys. \textbf{81} (2018) 026101.

\bibitem{CMS_TDR}
  CMS Collaboration,
  ``A MIP Timing Detector for the CMS Phase-2 Upgrade'',
  CERN-LHCC-2019-003.

\bibitem{ATLAS_TDR}
  ATLAS Collaboration,
  ``Technical Design Report: A High-Granularity Timing Detector for the ATLAS Phase-II Upgrade'',
  CERN-LHCC-2020-007.

\bibitem{WF2}
  F. Cenna et al.,
  Weightfield2: a fast simulator for silicon and diamond solid state detectors,
  Nucl. Instrum. Meth. A \textbf{796} (2015) 149--153.

\bibitem{Meroli2011}
  S. Meroli, D. Passeri, L. Servoli,
  ``Energy loss measurement for charged particles in very thin silicon layers'',
  JINST \textbf{6} (2011) P06013.


\bibitem{Riegler2025}
  W. Riegler,
  ``Time resolution limits in silicon sensors from Landau fluctuations and electronics noise'',
  Nucl. Instrum. Meth. A \textbf{1080} (2025) 170808,
  \url{https://doi.org/10.1016/j.nima.2025.170808}.

\bibitem{Leo1994}
  W.R. Leo,
  \textit{Techniques for Nuclear and Particle Physics Experiments},
  Springer-Verlag, 2nd edition, 1994.

\bibitem{Siviero2022}
  F. Siviero et al.,
  ``Optimization of the gain layer design of Ultra-Fast Silicon Detectors'',
  Nucl. Instrum. Meth. A \textbf{1033} (2022) 166739.


\end{thebibliography}
\end{document}